\documentclass[%
twocolumn,
nofootinbib,
amsmath, amssymb,
aps,prd,longbibliography
]{revtex4-2}

\usepackage{graphicx}
\usepackage{dcolumn}
\usepackage{bm}
\usepackage{amssymb}	
\usepackage{amsthm}	
\usepackage[sc]{mathpazo}
\usepackage{mathrsfs}
\usepackage{booktabs}
\usepackage{rotating}
\usepackage[flushleft]{threeparttable}
\usepackage{amsmath}
\usepackage{tikz,scalerel}
\usepackage{xcolor}
\usepackage[%
  colorlinks=true,
  urlcolor=blue,
  linkcolor=blue,
  citecolor=blue
]{hyperref}
\usepackage{etoolbox}
\usepackage{breqn}

\makeatletter
\let\cat@comma@active\@empty
\makeatother

\newcommand{\beq}{\begin{equation}}
\newcommand{\eeq}{\end{equation}}
\newcommand{\bea}{\begin{eqnarray}}
\newcommand{\eea}{\end{eqnarray}}
\newcommand{\Ur}{\mathrm{U}}
\newcommand{\Vr}{\mathrm{V}}

\newcommand{\epp}{\epsilon^{\prime}}

\begin{document}
\title{ 
The refractive index of the medium within the volume of deep inelastic $ep$ scattering:\\
I. Group velocity of interaction carriers
}

\author{B. B. Levchenko}%
\email{levchenko@www-hep.sinp.msu.ru }
\affiliation{%
 D.V. Skobeltsyn Institute of Nuclear Physics, M.V. Lomonosov Moscow State University,\\ 
 119991 Moscow, Russian Federation}%

\begin{abstract}
We propose a new method for measuring several characteristics of a fermion-boson medium formed during deep inelastic particle scattering. This method involves measuring the group and phase velocities of the accompanying wave process.
The phase and group velocities of a wave process are related by the Rayleigh differential condition. The refractive index of a medium is determined by the phase velocity, while the group velocity characterises the medium's dispersion properties.
In the case of deep inelastic $ep$ scattering under consideration, the wave process in an individual event is understood to be the exchange between an electron and a proton of an effective, non-perturbative mediator of interaction (${\cal B}$) with the quantum numbers of a neutral current.
In the first report, we present new data on the group velocity of interaction carriers  ${\cal B}$. 
 We extracted the  group velocity modulus values, $\Ur_{_{\cal B}}^*$, from the HERA collider data using a formula derived from modified Heisenberg uncertainty relations and applying the indirect measurement method.  
The HERA data demonstrate that in the selected
kinematic domain the  speed of the mediator ${\cal B}$ exceeds the speed of light $c$ in free space, $\Ur_{_{\cal B}}^* > c$. 
To compare with the obtained data on group velocity, we analyse a simplest model of dispersion  relation occurring due to the relativistic relationship between the energy, mass and momentum of a particle, supplemented by de Broglie relations. 

{\it Keywords: Lepton-Nucleon Scattering, Group Velocity, Refractive Index, Quantum Entanglement,  Superluminal  Velocity, HERA Collider }

\end{abstract}


\maketitle

\section{Introduction}
Since the introduction of group velocity as a new characteristic of wave processes about 150 years ago \cite{JWS1877a, JWS1877b}, this concept has repeatedly been subject to critical rethinking, refinement, and clarification of the conditions under which this value uniquely characterizes a wave process.  
During the study of the propagation of waves and wave impulses of different nature in deformable media with dispersion, about a dozen definitions of velocity have been proposed to date to characterize the process itself:
the phase velocity, the group velocity \cite{JWS1877a},
the signal velocity \cite{Som1914, Bri1914},
the  velocity of energy transport, which yields the rate of energy 
flow through a continuous wave  \cite{Bri1960},
 the centrivelocity \cite{Smi1970},
the correlation velocity \cite{Bl1977},
the pulse velocity \cite{Gur2001} etc.
The signal velocity (that propagates information) is different from the group velocity.
For the light propagation have to be added more two definitions,
the relativistic  velocity constant, the ratio of units velocity\cite{Smi1970}.
Each definition reflects new revealed features of the propagation of the wave process in the medium and the transfer of energy, momentum and information. 

For instance, in cases where a single infinite train of waves propagates in a homogeneous conservative, dispersive  media the mean energy flux velocity vector is equal to the group velocity  \cite{Hayes77}.
In these cases, the directions of propagation of the phase and group velocities coincide. At the same time, the group velocity is said to be positive.
However, there are dispersive  media in which the phase and group velocities in certain frequency regions are in opposite directions.
This is obtained in the so-called ``optical'' branches of the acoustic spectrum of the crystal lattice \cite{M1945}. 

Summing up his many years of research on wave processes conducted  in collaboration with Sommerfeld, Brillouin summarized the conditions under which the concept of group velocity applies 
as follows \cite{Bri1960}:
 ``It is apparent from the manner in which the group velocity was 
defined that this concept is wholly precise only when the wave packet is 
composed of elementary waves lying within an infinitely narrow region 
of the spectrum.'' 

It is also important that the special theory of relativity postulates the speed of light as the limiting speed of transmission of energy and information.
Under normal circumstances, the velocity of particles with nonzero mass does not exceed the speed of light. As de Broglie demonstrated, this velocity corresponds to the group velocity. This is postulated as a principle in the special theory of relativity.  However, this only applies under certain physical conditions.

Assuming the constancy of the speed of light in a vacuum, the special theory of relativity (STR) implicitly assumes the linearity of the theory. Indeed, Maxwell's equations (ME) are linear, ensuring the fulfilment of the principle of superposition of waves observed experimentally for linear media. Lorentz transformations ($\Lambda_0$) are a consequence of ME. Therefore, the relativistic invariance and covariance of theories within the framework of STR result from combining two postulates: 
$\Lambda_0\, \oplus\, (\sup \Ur=c)$. 
 However, the transformations $\Lambda_0$  are a consequence of a weak field approximation.
Therefore, for nonlinear equations, the transformations that preserve the theory's invariance and covariance should differ depending on the field's intensity, $\mathcal{F}$,
i.e. they should be given by the function  $\Lambda(\mathcal{F})$. In his book Lectures on Quantum Mechanics (Chapter 1: The Hamilton Method) \cite{Dir1964} , P. Dirac notes, ``one would like to take into account the possibility that Maxwell's equations are not accurately valid. When one goes to distances very close to the charges that are producing the fields, one may have to modify Maxwell's field theory so as to make it into a nonlinear electrodynamics.''

The foundations of nonlinear quantum electrodynamics  (QED) and strong field phenomena were laid by  works of Sauter  \cite{sau31},  Heisenberg and  Euler \cite{he36}, Schwinger \cite{schw51} and  Ritus \cite{rit75}.  
By analyzing the propagation of signals in nonlinear electrodynamics, Blokhintsev 
\cite{Blokh52, Blokh53} established that interactions in nonlinear field theory occur at speeds depending on the magnitude of the field. These speeds differ from the speed of light in a vacuum, $c$. In particular, they can be larger than $c$. In this case, the speed of interaction refers to the speed at which the wavefront carrying the interaction moves.

For QED, nonlinear effects only become significant at field intensities close to the critical value, $\mathcal{F} \sim \mathcal{F}_{cr}\sim 10^{16} V/cm$, 
 because the electromagnetic coupling constant is small, $\alpha_e\approx$1/137.
However, in the case of interactions between elementary particles, such as protons scattering by protons, quantum chromodynamics (QCD) takes over from QED. 
This is because the parameter of strong interactions, $\alpha_s $, is large (0.2–0.4), and QCD is nonlinear initially.

A study of photon propagation in gravitational field has also revealed a surprising phenomenon that
quantum corrections modify the characteristics of photon equation of motion in such a way that
photons may propagate superluminally  in the Schwarzschild \cite{Dru_Hath80},
Reissner-Nordstr$\ddot{\rm{o}}$m  \cite{DanSho94}, Kerr \cite{DanSho96}  and other backgrounds  \cite{DolgNov98}.

And last but certainly not least, let's recall  the phenomenon  of quantum entanglement.
 In 1935, while discussing the completeness of quantum mechanics, Einstein, Podolsky and Rosen (EPR) proposed a thought experiment \cite{EPR1935} which would later become known as the EPR paradox.
Despite numerous discussions, it took more than 30 years to develop a mathematical tool, Bell's theorem
 \cite{Bell1964}, which enabled the theoretical analysis to be expanded and an experimental study of the EPR paradox to begin.
Numerous experiments on the verification of Bell's inequalities and their generalisations (see, for example, \cite{Clauser1972, Aspect1982, Aspect2002}, \cite{Unspeak}) revealed the fundamental non-locality of quantum 
mechanics\footnote{The Nobel Prize in Physics  2022 was awarded to A. Aspect, J. Clauser and A. Zeilinger "for experiments with entangled photons, establishing the violation of Bell inequalities and pioneering quantum information science" \cite{NPP22}.}.
Nonlocality is understood as a deviation from the cause-and-effect sequence \cite{Weihs2002, Unspeak} 
  postulated by the special theory of relativity.
In entangled quantum systems, correlations between distant elements of the system are established almost instantaneously. Furthermore, no dependence of these correlations on the distance between parts of the system has been observed.
Experimental estimates of the speed of ``spooky action at a distance'' (the term coined by Einstein in 1947 to describe the phenomenon of quantum entanglement) yielded a lower limit of about $10^4$ times the speed of light for the propagation velocity of the correlation relationship  \cite{SpookyAct, Amato2023}.

Thus, a significant amount of theoretical examples and experimental evidences\footnote{
Notably,``fast light effects", when the group velocity exceeds the
speed of light $c$  have been demonstrated at an almost routine level 
for a variety of systems and media, see  references in \cite{Novikov:21}. }
 has been accumulated indicating that the superluminal, 
or ``faster-than-light" transfer of energy, momentum, and information is characteristic of some nonlinear 
processes or nonlinear media.

This article discusses the group and phase velocities of an elementary wave process in an individual scattering event.
The interaction between an electron and a proton is mediated by a  kind of elementary wave process 
(de Broglie waves) and is understood  to be  the exchange of an effective, non-perturbative mediator, ${\cal B}$, with quantum numbers corresponding to a neutral or charged current. In a perturbative approach to describing DIS  processes and for a  specific range of $Q^2$,
the mediator ${\cal B}$, to some approximation,  can be interpreted as one of the virtual gauge bosons, $\gamma^*, Z^0$, or $W^{\pm}$.

It is important to note the fundamental difference between an electromagnetic wave process or pulse with a finite spatial or temporal extent, and an elementary photon. Due to the innumerable number of photons with similar energies and momenta, as well as their bosonic nature, a collective state called the electromagnetic wave field is formed. In the case of a single scattering event, an intermediate wave process, a virtual mediator ${\cal B}$  (in particular, a virtual photon, $\gamma^*$), is produced for a short time. This particle carries quite a lot of ``information'' from the scattered particle (electron) to the scatter (proton). This type of ``information'' includes data on the  amount of transferred  energy-momentum, as well as on the charge\footnote{ Charge information is multifaceted and includes ``data'' about the electric, leptonic, baryonic, weak, and color charges of particles.} and spin state of particles.
 From a comparison of these two phenomena, it is clear that for a mediator ${\cal B}$, and due to the quantum uncertainty, 
all the velocity characteristics of wave processes listed above essentially combine into two: phase and group velocities.

In a recent publication \cite{bbl24}, based on modified Heisenberg's uncertainty relations (UR), a formula was derived to estimate a speed, $\Ur_{_{UR}}^*$,  
of an interaction carrier. We applied the theory of indirect measurements \cite{M72, Rab05} and the derived formula to estimate the speed of virtual photons from the HERA collider data \cite{DESY_15_039}.
The HERA data indicate that the speed of virtual photons  exceeds the speed of light $c$ in free space, $\Ur_{\gamma^*}^*>c$.  In Ref. \cite{bbl24} the term ``group velocity'' was used for
$\Ur^*$ and  one of the goals of this paper is to provide evidence to support the term used.

This paper is organized as follows. 
In the next section, we consider a three-dimensional version of Rayleigh's one-dimensional differential formula that connects  the group and phase velocities. 
Section 3 analyzes the simplest quantum mechanical model of the relativistic dispersion relation between the frequency and the wave vector of a particle, calculates the group and phase velocities, and verifies the three-dimensional version of Rayleigh's formula.
Section 4 begins by specifying  kinematic variables of  DIS processes. Then, it presents the analytical and graphical forms   of the above model group velocity, $\Ur_{_M}^*$.
In the same section, there is a formula that was recently derived from uncertainty relations. The formula 
$\Ur_{_{UR}}^*$ was used to estimate the norm of  group velocity of a virtual mediator  ${\cal B}$ 
from the DIS HERA data. The formula takes into account  resolutions of  detectors.
A comparative analysis was performed to determine how the expressions for 
$\Ur_{_M}^*$ and $\Ur_{_{UR}}^*$  
depend on the kinematic variables near points of the highest speed and close 
to the kinematic limits.

\section{3-D generalization of the 1-D Rayleigh differential condition}
In 1877, Lord Rayleigh derived a mathematical relationship between the phase velocity  
$\Vr$, the wavelength $\lambda$, and the group velocity $\Ur$ of propagating waves \cite{JWS1877a, JWS1877b}, 
\beq
\Ur=\Vr-\lambda\frac{\partial \Vr}{\partial \lambda}.
\label{R1}
\eeq
 ``The composition of two vibrations of nearly equal period'' was considered \cite{JWS1877a}.
In the one-dimensional problem, the wave velocities are determined according 
to  formulas
$\Vr= \omega/k$ (the phase velocity) and $\Ur =\partial \omega /\partial k$ (the group velocity),
with the wave number  $k=2\pi/\lambda$ and  $\omega$  the angular frequency. 
With such a definition of the velocities $\Vr$ and $\Ur$, Eq.  (\ref{R1}) turns into a mathematical identity. 
For a medium without dispersion, $\omega = v_0 k$, where $v_0$ is the phase velocity   in the medium in question.
In this case, the definitions of velocities and Eq. (1) imply that
 $\Vr=\Ur$ and $\Vr\Ur=v_0^2$.
Therefore, the group velocity is equal to the phase velocity.
Note that it was  Rayleigh who introduced the very concept (and name) of group velocity, one of the basic concepts of any wave theory, which plays such an important role in the theory of radio wave propagation, in optics, in acoustics, in geophysics, and in wave quantum mechanics.

When a wave propagates through a medium with a refractive index of $n$, the frequency ($\omega$), wave vector ($k$) and refractive index ($n$) are related  \cite{BornW}  by the equation $k= \omega n(\omega)/c$. Therefore, the phase velocity is directly  connected to the refractive index of the medium: $\Vr = c/n$. We will not discuss  this topic further, leaving it for a separate analysis.

 For the fully three dimensional case,
when the frequency depends on all three projections of the wave vector, the definition of the group velocity and the Rayleigh equation itself requires generalizations.
The group velocity vector in the coordinate system with the orthogonal unit vectors
$(\mathbf{e_1, e_2, e_3})$ is defined as follows,
\beq
\mathbf{U} = \frac{\partial \omega}{\partial k_1}\mathbf{e_1}
+\frac{\partial \omega}{\partial k_2}\mathbf{e_2}
+\frac{\partial \omega}{\partial k_3}\mathbf{e_3}.
\label{R}
\eeq
Each of the projections of the vector $ \mathbf {U} $ is related to the projections of the vector $ \mathbf {V} $ by an analogue of Eq. (\ref {R1}),
\beq
\Ur_i=\Vr_i+k_j\frac{\partial \Vr_i}{\partial k_j},
\label{Rgen}
\eeq
where summation is performed over repeated indices. Now let's substitute 
Eq. (\ref {Rgen}) to the square of the vector
 $\Ur^2 = \sum_{i=1}^{3}(\Ur_i)^2 $. Thus,
\begin{displaymath}
\Ur^2 = \sum_{i=1}^{3} \Big (V_i+k_j\frac{\partial \Vr_i}{\partial k_j} \Big)^2 .
\end{displaymath}
Opening the square brackets and  performing the replacement $k_j\partial \Vr_i/ \partial k_j =\Ur_i-\Vr_i$, we come to a generalization of the Rayleigh equation to the three-dimensional case,
\beq
2\mathbf{V}^2 + k_j\frac{\partial \mathbf{V}^2}{\partial k_j} 
-2({\mathbf{U}}\cdot \mathbf{V})=0.
\label{GRGgen}
\eeq
However, the dot product introduces the angle $\alpha$ between the vectors ${\mathbf{U}}$  
and ${\mathbf{V}}$, meaning they can point in different directions, including opposite ones.

In the next section, we  demonstrate the validity of Eq. (\ref{GRGgen}) using  a simple relativistic model of the dispersion relation, $\omega(\mathbf{k})$.

\section{Simple relativistic model of the dispersion relation}
In the probabilistic interpretation of quantum theory, a solution to the Schr$\ddot{\rm{o}}$dinger and/or Dirac equations for a free particle is represented by an infinitely long plane wave with a complex phase. Therefore, a particle is viewed as a non-localized state. Such plane waves describe the observed effects in repeated experiments, such as interference fringes in a double slit experiment.

On the other hand, quantum theory is based on the fact that photons are particles that are 
absorbed/radiated  by atoms as localized lumps of energy. 
Therefore, to avoid confusions, one may in fact use two different wave functions,
a localized individual single event wave function, and a nonlocalized plane wave function. 
Barut \cite{Barut90} constructed an example of the first type of wave function. It is a model of quantum particles that incorporates the basic relations of particle-wave duality and the relativistic relation between energy, momentum, and mass. It represents a three-dimensional, localized, finite field energy soliton-like oscillating lump that does not spread. 
When dealing with a virtual mediator  ${\cal B}$, we will keep this localized solution, 
found by Barut, in mind as a visual image of it. 
This is also justified by  the Brillouin condition for the applicability of the concept of group velocity and
the fact that Barut constructed  a localized tachyonic solution with a group velocity greater than $c$ \cite{Barut1993}.

Let us denote by $q$ a  virtual  ${\cal B}$ energy-momentum
4-vector with components  $(q_0/c, \mathbf{q})$.
The energy and momentum of a particle are related to its frequency, $\omega$, and wave vector, 
$\mathbf{k}(k_x, k_y, k_z)$, via de Broglie relations:
$q_0^2 = (\hbar \omega)^2$ and ${\mathbf{q}=\hbar \mathbf{k}}$.
 The frequency, $\omega$, is a function of $\mathbf{k}$, and is uniquely determined by the nature of the wave. For an electromagnetic wave in a vacuum, it is simply given by the equation:
$\omega^2=c^2( k_x^2+ k_y^2+ k_z^2)$.

Consider the relativistic relation between energy, momentum and mass as a simplest model of the dispersion relation:
\beq
q_0^2/c^2 = (m^*c)^2 +\sum q_i^2,
\label{e1}
\eeq
from which follows
\beq
\omega^2 =  \omega_0^2+c^2\sum k_i^2,
\label{e2}
\eeq
where $\omega_0^2=(m^*c^2/\hbar)^2$  
and $q^2c^2=(m^*c^2)^2$.

Then, in accordance with Eq. (\ref{R}), the square of 
 group velocity of the  virual  ${\cal B}$ is
\beq
(\Ur_{_M}^*)^2 =  \Big (\frac{\partial\omega}{\partial k_x}\Big )^2 +
\Big (\frac{\partial\omega}{\partial k_y}\Big )^2 +
\Big (\frac{\partial\omega}{\partial k_z}\Big )^2,
\label{e3}
\eeq
where, for instance,
\beq
\frac{\partial\omega}{\partial k_x}= \frac{c^2q_x}{|q_0|}.
\eeq
Thus,
\beq
\lvert \Ur_{_M}^*\rvert =c^2\frac{\lvert \mathbf{q} \rvert}{|q_0|}.
\label{e4}
\eeq

From the definition of the phase velocity and Eq. (\ref{e2}), with
account $q^2c^2=q_0^2-c^2\mathbf{q}^2$, one get
\beq
\lvert \Vr_{_M}^*\rvert=\frac{\omega}{k}=\frac{\sqrt{(m^*c^2)^2  + c^2\hbar^2 k^2}}{\hbar k}=\frac{|q_0|}{\lvert \mathbf{q} \rvert}.
\label{e5}
\eeq
It is interesting to note that $\lvert \Vr_{_M}^*\rvert \lvert \Ur_{_M}^*\rvert=c^2$, just as in the case of a dispersion-free medium or a transparent plasma medium containing free charge carriers (the ionosphere).

We can now substitute the expressions for $\Vr_{_M}^*$ and $\Vr_{_M}^*\Ur_{_M}^*$  into equation (\ref{GRGgen}), which reduces to the following condition: $2c^2(1-\cos\alpha)=0.$
Thus, relation (\ref{GRGgen}) is satisfied if $\alpha=0,$ meaning the directions of velocity vectors 
$\mathbf{V_{_M}^*}$ and $\mathbf{U_{_M}^*}$ coincide. This is true for the dispersion model  (\ref{e2}).

After specifying the kinematic variables of  DIS processes in the next section, we will present a graph 
of the function (\ref{e4}).

\section{DIS kinematcs, the model predictions and the HERA data for  group velocities }
As mentioned above, one source of virtual bosons is the hard scattering of charged leptons (e.g., electrons and positrons) on protons. Therefore, let's briefly remind the definitions of some kinematic variables that describe the particle scattering process and that are used later.
For  lepton-proton scattering at the HERA collider let's
denote the components of the incoming lepton and proton 4-momenta
as follows
$${\cal K}\ =\ (\epsilon/c, 0,0,-\ell),\ \ \ \ {\cal P}\ =\ (E/c, 0,0,\,P)\,,$$
and also mark the scattered lepton with a prime\footnote{The  right-handed 
Cartesian coordinate system has its origin  at the nominal interaction point of two beams, 
the $Z$ axis pointing in the proton beam direction, and the X axis pointing toward 
the center of HERA \cite{DESY_15_039}.}. 
The scattered lepton makes an angle $\theta$ with incoming lepton.
In these notations we
derive below an expression for the virtual boson energy\footnote{This 
formula plays a vital role in revealing the  properties of a virtual  ${\cal B}$. }
\beq
q_0\,=\,\epsilon\,-\,\epp\,=\,%
\frac{c^2(\ell-xP)Q^2}{2x(\ell E+\epsilon P)}\simeq c(\ell-xP)y\,,
\label{kinema}
\eeq
which relates the scattered lepton energy to negative
four-momentum-transfer squared, $Q^2$,  components
of the incoming  lepton and proton four-momenta  and 
$x=Q^2/2(q{\cal P})$, the Bjorken variable, $0<x<1$. Here 
$y=(q{\cal P})/({\cal KP})$, $0<y<1$. The energy-momentum four-vector of a mediator  ${\cal B}$  
 is given by $q ={\cal K}\,-\,{\cal K}^{\prime}$, 
or in components, $q = (q_0/c,\mathbf{ q})$.

One need to exclude the term $\ell^{\prime}\cos\theta$ from the both expressions below:
\bea
q^2\,&=&\,-Q^2\,=\,({\cal K}\,-\,{\cal K}^{\prime})^2 \nonumber  \nonumber \\
&=&2\,\epsilon q_0/c^2\, -2\ell^2\, +
\,2\,\ell \ell^{\prime}\,\cos \theta \nonumber 
\eea
and
$$ (q{\cal P})\,=\,\frac{Q^2}{2\,x}\,=\,Eq_0/c^2\,+\,P\ell\,
-\,P\ell^{\prime}\cos \theta\,.$$
In the result, we receive the first part of Eq.\,(\ref{kinema}).
In the massless approximation\footnote{The massless approximation 
corresponds to replacing  $s-m_e^2c^4 -m_p^2c^2$ with $s$.}
with use of the relations $c^2Q^2=xys$ and $s\,=\,4\,\ell Pc^2$
one gets 
$$q_0\,=\,c(\ell\,-\,x\,P)\,y\,.$$

\begin{figure}[t]
\centering
\includegraphics[height=0.46\textwidth,width=0.48\textwidth]{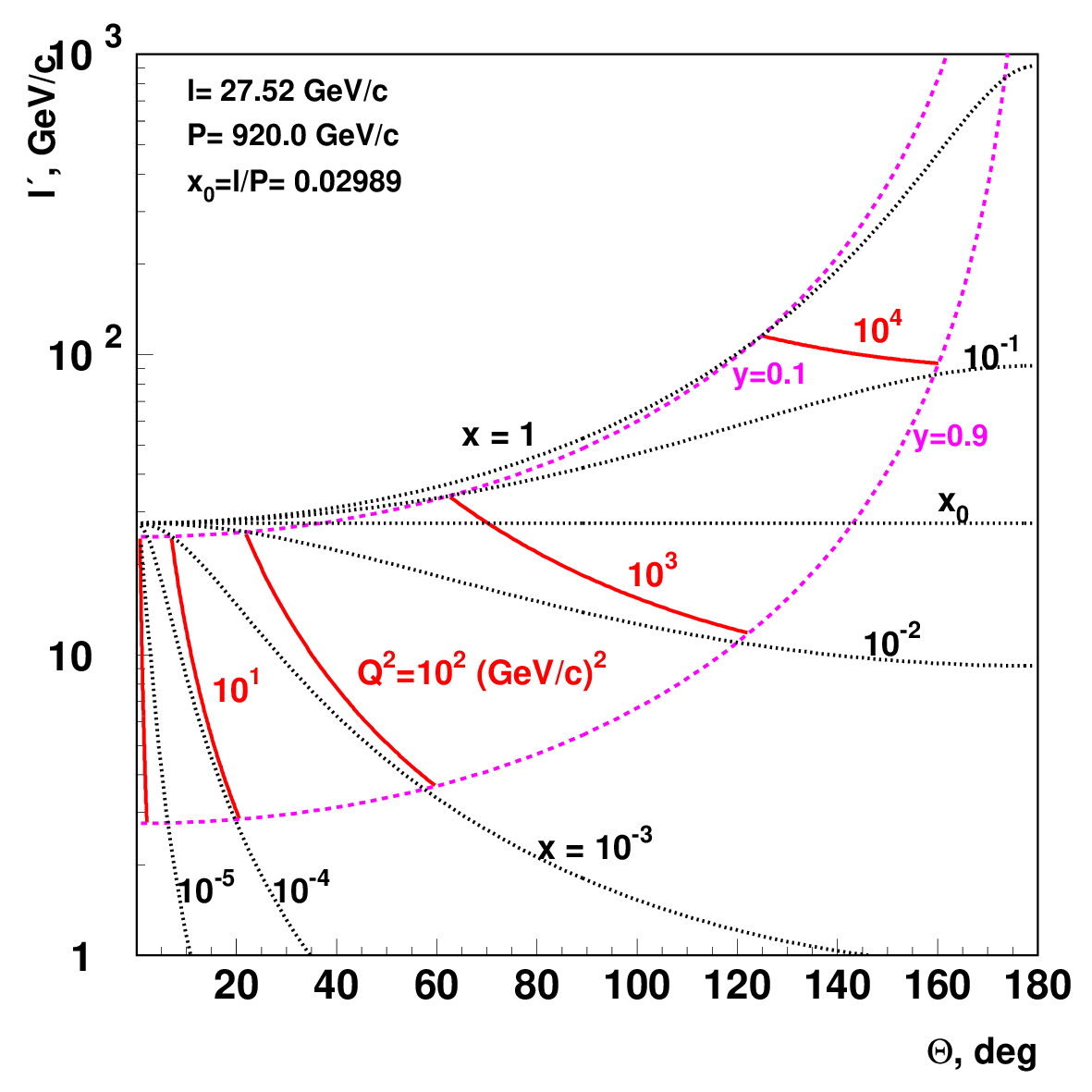}
\caption{Isolines of the kinematic variables $x$ (short-dashed, black), $Q^2$ (full, red), 
and  $y$ (long-dashed, magenda) are shown in the ($\theta$, $\ell^{'}$) phase space for 
electron-proton scattering at the HERA collider. The polar angle, $\theta$,  of the scattered lepton  is defined 
with respect to the direction of the incoming lepton. The  momentum of the scattered lepton 
 is $\ell^{'}$. The line marked $x_0$ corresponds to space-like virtual bosons.}
\label{fig1} 
\end{figure}

From Eq. (\ref{kinema}) it is easy to find that  along the line $x=\ell/P=x_{_0}$ a
virtual  ${\cal B}$ is purely space-like, $q_0=0$, already in the HERA laboratory frame, regardless of  
 $Q^2$ or $y$ values. 
Figure \ref{fig1} clearly shows this feature of the $x_0$ line.
Figure  \ref{fig1} shows a panoramic view of electron-proton scattering kinematics, demonstrating how the momentum of the scattered lepton depends on the variables 
$x$, $Q^2$ and $y$ when plotted as a function of its scattering angle.
Additionally, it can be seen from Eq. (\ref{kinema}) that, for $x < x_0$, the boson energy $q_0$ is positive, meaning the electron lost energy in the scattering process.
However, when $x$ is greater than $x_0$, the boson energy is negative, indicating the transfer of energy from the proton to the electron.

\begin{figure}[t]
\centering
\includegraphics[height=0.49\textwidth,width=0.49\textwidth]{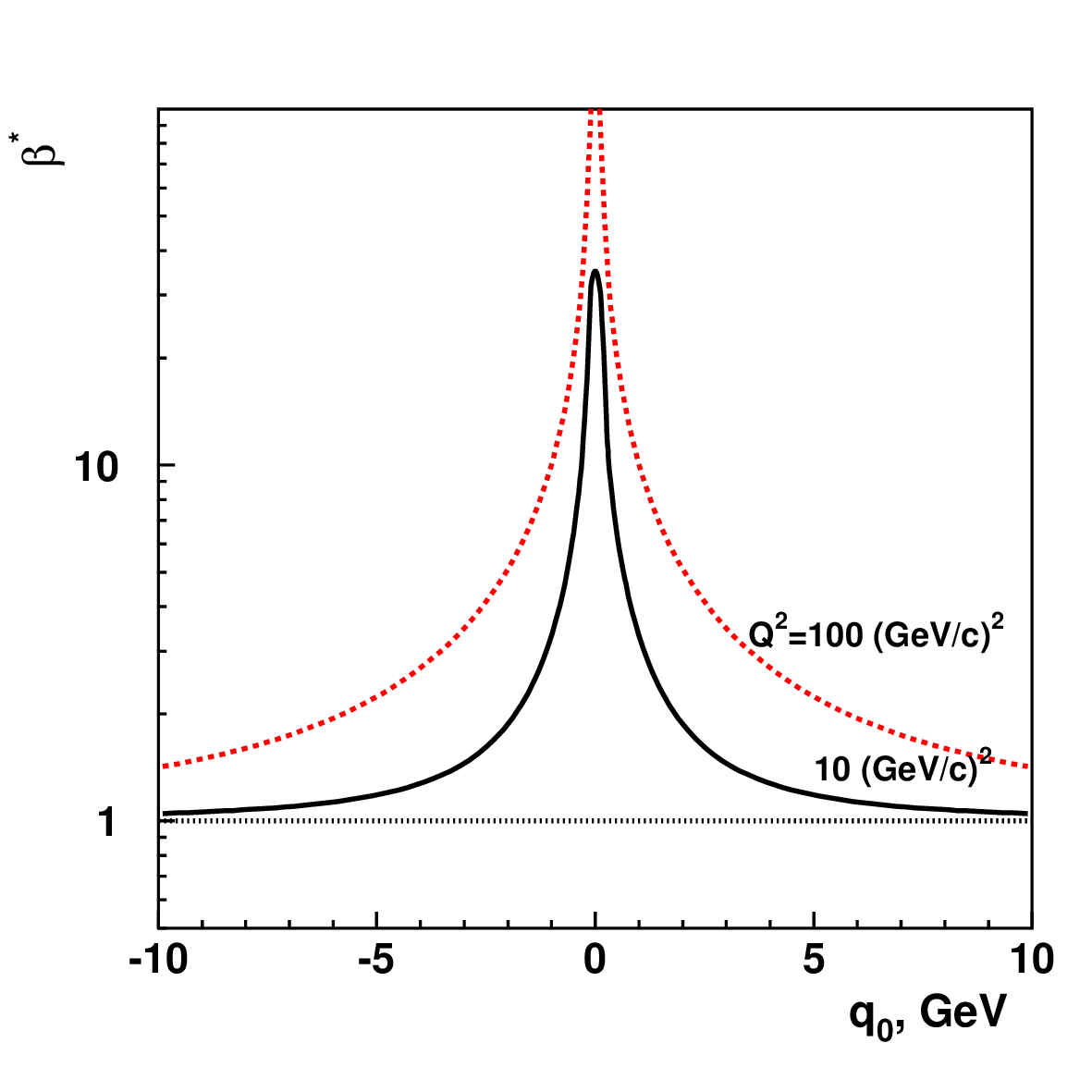}
\caption{The normalized virtual boson speed accoding to the model based on
the  relativistic dispersion relation (\ref{e2}) and the HERA collider kinematics.
}
\label{fig2} 
\end{figure}

Now that we have the definitions and ranges of the kinematic variables in DIS processes, we can plot 
the dependence of $ \Ur_{_M}^*$ on the energy of the virtual boson, $q_0$.
 Let's  normalize the speed of the virtual particle to that of a real photon in  vacuum,  $\beta_{_M}^*= \Ur_{_M}^*/c$.
 Using the kinematic variables introduced above, transform the expression  (\ref{e4})  to the following form,
\beq
\beta_{_M}^*(q_0)= \sqrt{1+\frac{c^2Q^2}{q_0^2}}.
\label{beta_m}
\eeq
Figure  \ref{fig2} shows   the dependence of  $\beta_{_M}^*$   on $q_0$   at two values of $Q^2$.
 According to Eq. (\ref{beta_m}) and the definition of  $q_0$, as 
$q_0$ approaches 0, $\beta_{_M}^*$ tends to infinity. The plots in Fig.  \ref{fig2} only able 
to demonstrate this tendency.
However, it should be noted that, according to Eq. ({\ref{kinema}), the point $q_0=0$ can 
be reached via two paths: 
when $x$ approaches $x_0$ and/or when $y$ approaches the kinematic limit of $y=0$.
 Thus, within the framework of the dispersion model under consideration, it was found 
that $\lvert \Ur_{_M}^*\rvert > c$, $\lvert \Vr_{_M}^*\rvert <c$ and  
$\lvert \Vr_{_M}^*\rvert \lvert \Ur_{_M}^*\rvert=c^2$ .
At the same time, according to Eq. (\ref{beta_m}), when $Q^2=0\,  {\rm GeV^2/c^2}$,
the particle virtuality is equal to zero, $\lvert \Ur_{_M}^*\rvert=\lvert \Vr_{_M}^*\rvert =c$.

In the recent publication \cite{bbl24} based on a completely different concept
we derive a formula for estimating the speed $\Ur_{_{UR}}^*$ of a virtual particle in indirect measurements,
\beq
\vert \Ur_{_{UR}}^* \vert
=c\sqrt{\frac{(\Delta q_0)^2}{(c\Delta q)^2}},
\label{Vel_lb}
\eeq
where $\Delta q = \lvert \mathbf{\Delta q}\rvert$.
We applied the theory of indirect measurements to calculate the 
quantities  $(\Delta q_0)^2$  and $ (c\Delta q)^2$. 
Formula (\ref{Vel_lb})  allows to estimate the module of the group velocity of virtual bosons 
from the DIS HERA data.
In a generalised sense, formula  (\ref{Vel_lb})  is the mathematical analogue of the definition of group velocity (\ref{e3}), but takes into account the inaccuracies of the measurement process (resolutions of measuring equipment).

 The quantities   $(\Delta q_0)^2$  and $ (c\Delta q)^2$  depend on the kinematic variables  $x$, $y$, $Q^2$ and   uncertainties of their measurements by the following chain of relations  \cite{bbl24},
\bea
(c\Delta q)^2 &=& \Big ( \frac{ q_0 }{ c|\vec{q}|} \Big )^2 (\Delta q_0)^2 
+ \frac{(c^2\Delta Q^2)^2}{4(c\vec{q})^2}  ,\label{un_p} \\
(\Delta q_0)^2 &=& c^2P^2y^2 (\Delta x)^2
+\,\frac{q_0^2}{y^2}(\Delta y)^2, \label{un_e} 
\eea

\bea
(c^2\Delta Q^2)^2 &=& \frac{4 c^2Q^2}{1-y} (c\Delta p_t)^2
+  \Big (\frac{c^2 Q^2}{1- y } \Big )^2(\Delta y)^2,\label{un_q2} \\
(\Delta x)^2 &=& \frac{4 x^2}{(1-y)c^2 Q^2 } (c\Delta p_t)^2\nonumber \\
&& +\,x^2\Bigg [\frac{1}{(1- y)^2 }+\frac{1}{y^2 } \Bigg ](\Delta y)^2.\label{un_xb}
\eea
The chain of these relations is closed when entering  the resolution of the ZEUS central tracking  detector $\sigma(p_t)/p_t$ and the energy resolution of the ZEUS uranium calorimeter  $\sigma(\Sigma_e)/\Sigma_e$  \cite{zeus_ctd, zeus_cal}, which are denoted as $c\Delta p_t$ and $\Delta y$, respectively, in Eqs (\ref{un_p})-(\ref{un_xb}).
As input to Eqs (\ref{Vel_lb})-(\ref{un_xb}), we use the combined data from the H1 and 
ZEUS experiments on deep inelastic $ ep $ scattering at the HERA collider \cite{DESY_15_039}.
%
\begin{figure}[t]
\centering
\includegraphics[height=0.46\textwidth,width=0.48\textwidth]{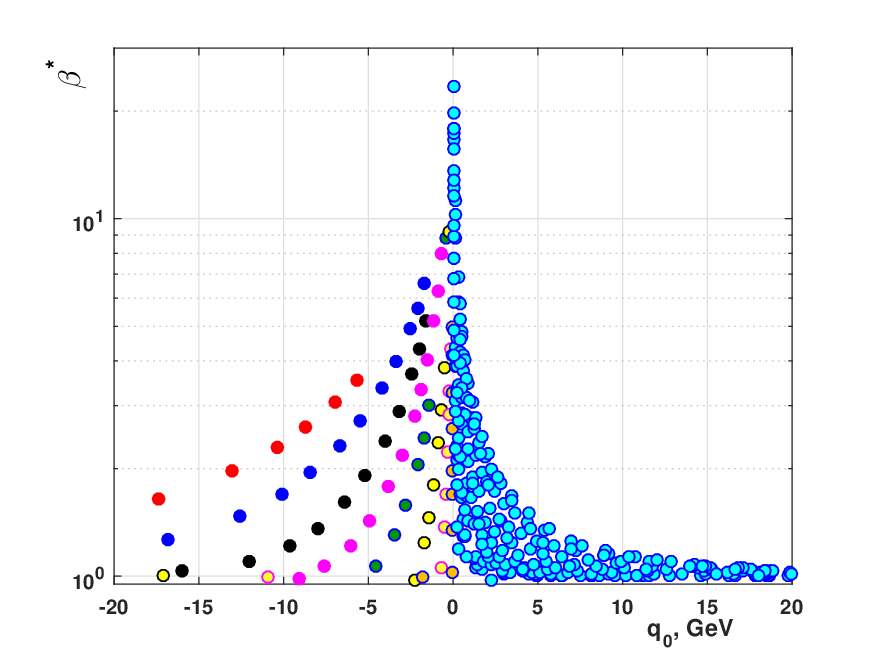}
\caption{The normalized speed of    ${\cal B}$, $\beta^* = \vert \Ur_{\cal B}^* \vert/c$, 
as a function of the   energy $q_0$ extracted  with the use of Eqs. (\ref{Vel_lb})-(\ref{un_xb})
from  the combined HERA II   data  
for neutral current  $e^+p$  deep inelastic scattering events at the center-of-mass energy 
 $\sqrt{s}\approx 318$ GeV   \cite{DESY_15_039}, Table 10.
The color of the markers corresponds to different values of the  variable $x$, as 
described in Table 1.}
\label{fig3} 
\end{figure}

To correctly correlate the obtained result with the type of mediator of interaction, bear in mind that, at low $Q^2 \le 100\, {\rm GeV^2/c^2}$, neutral current interaction cross sections are dominated by virtual photon exchange. However, for much larger values of $Q^2$, the exchange of a virtual $Z$ boson dominates neutral current processes in $ep$ scattering.

Figure~\ref{fig3} shows the  speed of virtual  ${\cal B}$ normalized to the speed of real photons   as a function of  $q_0$  at different $x$-values. 
Data points are grouped into strips with same  values of  $\bar{x}$ and marked with different colors (see  caption to Table 1). Such structuring reflects the kinematic relationship between the variables $x$, $y$ and $Q^2$, and the procedure for combining data from two different experiments by translation onto common grids \cite{DESY_15_039}.

\begin{table}[!ht]
\renewcommand{\arraystretch}{1.25}
\renewcommand{\tabcolsep}{3pt}
\begin{center}
\small  
\begin{tabular}{l|c|c|c|c|c}\hline
   Color          & $\bar{x}$       &$\beta^*_m$ & $Q^2$,  ${\rm GeV^2/c^2}$     & $y$        & $q_0$, GeV  \\ \hline
Red                & 0.65     &3.530            &650.0                            & 0.0099 & -5.633                \\ 
Blue                & 0.40     &6.573           &200.0                         & 0.0049   & -1.681               \\ 
Black              & 0.25      &5.177          & 200.0                        & 0.0079  & -1.60                  \\
Magenda        & 0.18     &7.990           & 90.0                          & 0.0049  & -0.682                  \\
Green            &  0.13     & 8.846        & 60.0                          & 0.0046  & -0.420                 \\
Yellow            &  0.08     & 9.16        & 35.0                          & 0.004  & -0.20                \\
Yellow-1         &  0.05     & 4.3        & 45.0                          & 0.009  & -0.16                \\
Yellow-2        &  0.032     & 5.0       & 22.0                          & 0.007  & -0.013                 \\
Cyan               & $\leq$ 0.013    & 23.522    & 1.5                           & 0.0011  & 0.0178               \\ \hline
\end{tabular}\label{tab:table1}
\caption{ 
The colour of the markers in Fig. \ref{fig3} is matched to the value of $\bar{x}$ in the strip.
 The largest values  of $\beta^*=\beta^*_m$ for a given value of $\bar{x}$ and their accompanying 
values of variables  $Q^2$, $y$ and $q_0$ are given too. 
}
\end{center}
\end{table}

The results presented in Fig.~\ref{fig3} show that the speed of  ${\cal B}$  
in DIS events of $ep$ scattering at  $ Q^2 > 0 \,  {\rm GeV^2/c^2}$ exceeds the speed of light $c$ in free space, 
$\beta^*>1$. 
Along with this, as can  be verified using Eq. (\ref{Vel_lb}) and relations 
(\ref{un_p}) - (\ref{un_xb}) \cite{bbl24}, as $Q^2$ approaches zero, there is a transition of the speed to that of a real photon, $\Ur_{\cal B}^*=c$.
As can be seen in Fig.  \ref{fig3} and Table 1, events  involving the fastest  superluminal virtual bosons  are localised  between the isolines $x=1$ and $y=0.1$ (see Fig.  \ref{fig1}) ,  
and the closer $q_0$ is to zero, the higher the value of $\beta^*$.
The largest number of events with superluminal  $\gamma^*$ has been found to be created 
in the phase space region at $q_0 > 0$ GeV and $x < 0.01$ (see cyan marker). Recall that if $q_0 > 0$ GeV,
this means that an electron is scattering in the forward direction, within the $\theta <\pi/2$ domain 
(see  Fig. 1).

Comparing Figs  \ref{fig2} and \ref{fig3} reveals a rough similarity in the way the function 
$\beta^*$ depends on the  energy $q_0$.  In both cases, the point $q_0=0$ GeV is special, 
as this is where  functions  $\beta^*(q_0)$ takes their maximum values. However, there are also 
many differences between the figures. For instance, the curve $\beta_{_M}^*(q_0)$ is symmetric 
with respect to the point $q_0=0$ GeV in Fig. 2, whereas there is no such symmetry in Fig. 3. This can be explained by the asymmetry of the initial electron and proton beam momenta. As a result, the surface $x_0=\ell/P$ divides the DIS process phase space into two significantly different volumes.

To be more specific, let's compare the behavior of  functions (\ref{beta_m}) 
and (\ref{Vel_lb}) near the point $q_0=0$ GeV. As mentioned previously, there are two ways to approach the point $q_0=0$ GeV: $x\to x_0$ and $y\to 0$. Therefore, we use $\delta$ to denote a small deviation from 
the point $x_0$: $x = x_0 -\delta$. 
For the model   (\ref{e2}), the expression (\ref{beta_m})  transforms to
\beq
\beta_{_M}^*(\delta, y)_{ q_0\to 0}\sim \sqrt{\frac{4x_0^2}{\delta^2y}},
\eeq}
which is singular in both $\delta$ and $y$.

The formula (\ref{Vel_lb}) requires slightly more complex algebraic transformations.
We are interested in the limit as $y\to 0$, so in the expressions (\ref{un_p})-(\ref{un_xb}), 
we replace $1-y$ by unit. Now substitute (\ref{un_xb}) into (\ref{un_e}) and (\ref{un_q2}) and (\ref{un_e})
into (\ref{un_p}). As a result, one gets
\begin{widetext}
\beq
\beta_{_{UR}}^{*^2}\approx \frac{4E^2[E^2\delta^2 y^2+s x y][s(\delta^2+x^2+x^2y^2)(\Delta y)^2+4xy(c\Delta p_t)^2]}
{s[(sxy)^2+4E^4 \delta^2 y^2(\delta^2+x^2+x^2y^2)](\Delta y)^2+4xy(s^2+4E^4 \delta^2 y^2)(c\Delta p_t)^2}.
\label{approx2}
\eeq
\end{widetext}
Let us keep the terms with $y$ and $\delta$ to powers of one or less in the numerator and take the quadratic powers in the denominator into account. In this case, Eq.  (\ref{approx2}) is simplified,
\beq
\beta_{_{UR}}^{*^2}\approx \frac{4E^2x[s x(\Delta y)^2+4y(c\Delta p_t)^2]}
{s[sxy(\Delta y)^2+4(c\Delta p_t)^2]},
\label{approx3}
\eeq
and we find an expression for estimating the group velocity of virtual bosons near the point $q_0=0$ GeV, accounting for the hardware resolution. 
There is no longer any dependence on the small parameter $\delta$.
This reveals the fundamental difference between a more general expression (\ref{Vel_lb}) and a simple model (\ref{beta_m}).
The functions $(\Delta y)^2$ and $(c\Delta p_t)^2$ parameterise the energy and momentum resolution of the ZEUS detector, respectively, and are given by the following expressions:
$(\Delta y)^2 =(0.18)^2(1-y)/2\epsilon,$
$(c\Delta p_t)^2= [0.0058(cp_t)^2]^2 + [0.0065 (cp_t)]^2 +[0.0014]^2$, 
with $(cp_t)^2$=sxy(1-y) \cite{zeus_ctd, zeus_cal}.
Interestingly, the singular behaviour of $\beta^*$ near  $q_0=0$ GeV is only realised in the abstract scenario of ideal momentum resolution, when $\Delta p_t=0$ GeV/c,
\beq
\beta_{_{UR}}^*(x, y)_{ q_0\to 0}\sim \sqrt{ \frac{4xE^2}{sy}}.
\eeq
However, with a finite  momentum resolution, even along  the kinematic limit line $y=0$, $\beta^*$ may reach   large but finite values.
For example, by setting $y = 0$ in Eq. (\ref{approx3}) and taking into account the values of 
$(\Delta y)^2$ and $(c\Delta p_t)^2$, we find that
\beq
{c\beta_{_{UR}}^*}( x=x_0,y=0) \approx  \sqrt{\frac{x_0^2E^2(0.18)^2}{2\epsilon (0.0014)^2}}
\approx 477c.
\label{approx4}
\eeq
It should be noted here specifically that the limit transition for only  one of the variables, $y\to 0$, does not take into account the interdependence between the variables $Q^2$, $x$ and $y$, which is present in the data. For this reason, the obtained limit value (\ref{approx4}) is significantly overestimated compared to the values shown in Fig. \ref{fig3}.

\begin{figure}[t]
\centering
\includegraphics[height=0.46\textwidth,width=0.48\textwidth]{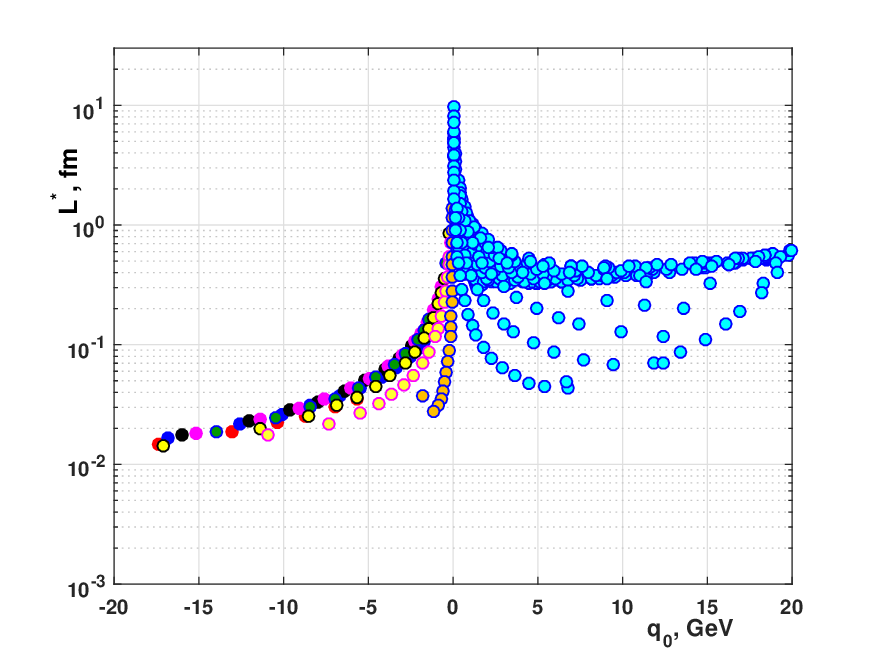}
\caption{The virtual ${\cal B}$  absorption mean free path $L^*=c\beta^* \Delta t$ (in femtometers)
as a function of the  energy $q_0$,  
extracted  with the use of Eqs (\ref{Vel_lb})-(\ref{un_xb})
from  the combined HERA II   data   for neutral current  $e^+p$  deep inelastic scattering 
events at the center-of-mass energy 
 $\sqrt{s}\approx 318$ GeV   \cite{DESY_15_039}, Table 10.
The color of the markers is the same as in Fig. \ref{fig3} and 
corresponds to different values of the  variable $x$.}
\label{fig4} 
\end{figure}

Using the speed of  virtual bosons and the relations given in equations  (\ref{un_p}) to (\ref{un_xb}), we can estimate  the absorption mean free path of a boson, $L^*$. This, in turn, provides an estimate of the magnitude of the impact parameters with which electrons interact with protons. Using Eq.  (\ref{un_e}) and the energy-time uncertainty relation, $\Delta q_0 \Delta t \approx \hbar$, we  derive $L^*=c\beta^* \Delta t$. 
The resulting distribution of $L^*$ is shown in Fig. \ref{fig4}. As can be seen from this distribution, the majority of interactions occur with $L^*\leq 1$ fm. However, as $q_0$ approaches zero, 
the absorption mean free path of a boson can be  large,
with $L^*\ge 10$ fm, which is much larger than the size of a proton.

\begin{figure}[t]
\centering
\includegraphics[height=0.46\textwidth,width=0.48\textwidth]{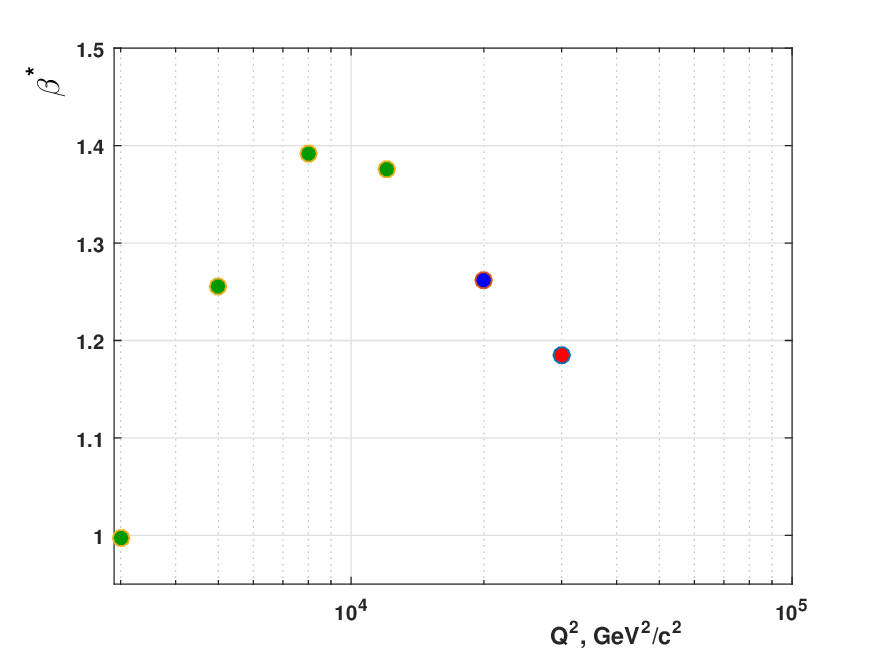}
\caption{The normalized  speed of ${\cal B}$, $\beta^* = \vert \Ur_{\cal B}^* \vert/c$, 
as a function of  $Q^2$ extracted  with the use of Eqs (\ref{Vel_lb})-(\ref{un_xb})
from  the combined HERA II   data  
for neutral current  $e^+p$  deep inelastic scattering events at the center-of-mass energy 
 $\sqrt{s}\approx 318$ GeV   \cite{DESY_15_039}, Table 10.
The color  of  markers corresponds to different  $y$-windows: 
the red dot has $y\in[0.7,0.75]$, the blue dot has $y\in[0.75,0.8]$, and the green dots have $y\in[0.9,0.95]$.}
\label{fig5} 
\end{figure}

In conclusion of this section, we make the following note. 
When analyzing real data, only events that satisfy a some set of conditions are selected in order to suppress background events. One of these conditions is selection of the window $y_{min} \le y \le y_{max}$. Consequently, only events located at a certain distance from the kinematic boundaries are analysed.  Thus, if the criteria for selecting events are too strict, the phenomenon under discussion may be overlooked.
Examining formulas (\ref{un_p}) to (\ref{un_xb}), 
it may appear that the limit $y\to 1$  leads to singular behaviour of the function $\beta^*$. 
However, this is  not the case.  
Indeed, when we substitute formulas (\ref{un_p}) - (\ref{un_xb}) into Eq. (\ref{Vel_lb}) and 
take the limit as $y$ approaches 1, we obtain
 \beq
{\beta}_{_{y\to 1}}^{*2}\approx \frac{q^2_0+ c^2Q^2}{q^2_0+ 4\epsilon^2}
\approx \frac{c^2Q^2}{4\epsilon^2}.
\eeq 
Thus, in the considered limit, the speed of a virtual particle is independent of hardware factors, exceeding the speed of light $c$ only at $c^2Q^2 > 4\epsilon^2$. For an electron beam with the energy of 27.5 GeV at HERA, this occurs when $Q^2 > 3025\, {\rm GeV^2/c^2}$. 
Figure \ref{fig5}  illustrates the accuracy of this estimate by showing the results for the region of $Q^2 > 10^3\, {\rm GeV^2/c^2}$ and $y > 0.7$. These results were obtained 
from the HERA data with using  the full set of Eqs. (\ref{Vel_lb})-(\ref{un_xb}). 
At $Q^2 > 10^3 \,{\rm GeV^2/c^2}$, the value of  $\beta^*$ is expected to increase. 
However, due to the small number of events in this $Q^2$ region, it is not possible to identify a regular subgroup of data points with  $y\to $1. Nonetheless, there is a growing trend of 
$\beta^*$ with increasing $y$. In Fig. \ref{fig5}, the red dot has $y\in[0.7,0.75]$, the blue dot has $y\in[0.75,0.8]$, and the green dots have $y\in[0.9,0.95]$.
As a reminder, with these $Q^2$, the exchange of a virtual $Z$ boson dominates the neutral current cross section in $ep$ scattering.

This most likely explains the significant difference in magnitude between the highest speeds measured in regions of low and high $Q^2$.

%


\section{Conclusions and outlook}
Extensive data on proton structural functions and extracted quark and gluon density distributions indicate that a fermion-boson medium (FBM) is formed during deep inelastic $ep$ interactions in the collision zone.
The interaction between an electron and a proton (or a nucleus) is understood to be the exchange of an effective, non-perturbative mediator, ${\cal B}$, with quantum numbers corresponding to a neutral or charged current.
The virtual ${\cal B}$ propagates    as an elementary wave process   in this medium. 
Access to additional characteristics of this medium, such as the refractive index, density, and plasma frequency, is possible if the phase velocity of the wave process is known.
The phase and group velocities are connected by the Rayleigh relation (\ref{GRGgen}). 
We have developed and tested a new tool, the formula (\ref{Vel_lb}), for extracting the modulus of group velocity  from experimental data. 
This formula is the three-dimensional mathematical equivalent of Rayleigh's definition of group velocity, which takes into account measurement inaccuracies. Thus, in the first stage of our program, 
we measured in each event the group velocity modulus of the  interaction carrier ${\cal B}$ across a wide range of kinematic variables. The group velocities were determined as a function of $Q^2$ and $x$ and presented as a function of $q_0$.
 The HERA data demonstrate that in the selected kinematic domain the speed of ${\cal B}$
exceeds  the speed of light $c$ in free space, $\Ur_{\cal B}^* > c$  \footnote{We also note that  in plasma media, there are many examples of group velocities being larger than $c$ \cite{Pec2013}.}. 
The speed of  ${\cal B}$ increases sharply as the ${\cal B}$'s energy, $q_0$, approaches zero.
To compare with the obtained data on group velocity, we analyse a simplest
model of dispersion relations occurring due to the relativistic relationship between the energy, mass
and momentum of a particle, supplemented by de Broglie relations.
Within this model and the kinematics of deep inelastic scattering, the group ($\Ur_{_M}^*$)  and phase ($\Vr_{_M}^*$) velocities of a wave process are calculated using Rayleigh's definitions. It is found that the velocities of $\Ur_{_M}^*$ and $\Vr_{_M}^*$ are related by the well-known equation $\Ur_{_M}^*\Vr_{_M}^*= c^2$.
Despite their approximate external similarity in terms of energy, the velocity distributions of bosons in the model and the data revealed significant differences in their kinematic structure near the singular point of $q_0 = 0$ GeV.

In the next report we will discuss the optical properties of the medium in the particle interaction zone.

\begin{acknowledgments}
The author is grateful to E. Boos, L. Dudko, J. Gao, A. Geizer, E. Oborneva, I. Volobuev 
and other colleagues for discussions and comments.\\

The study was conducted under the state assignment of Lomonosov Moscow State University.

\end{acknowledgments}


\bibliography{refracind1V2}

\end{document}